\font\bbb=msbm10                                                   %%%
%\def\bbb{\bf}                                                     %%%
%%%                                                                %%%
%%%                                                                %%%
%%%%%%%%%%%%%%%%%%%%%%%%%%%%%%%%%%%%%%%%%%%%%%%%%%%%%%%%%%%%%%%%%%%%%%
%%%%%%%%%%%%%%%%%%%%%%%%%%%%%%%%%%%%%%%%%%%%%%%%%%%%%%%%%%%%%%%%%%%%%%

\def\C{\hbox{\bbb C}}
\def\N{\hbox{\bbb N}}
\def\R{\hbox{\bbb R}}
\def\Z{\hbox{\bbb Z}}

\def\CMP{{\sl Commun.\ Math.\ Phys.}}

\def\CS{{\sl Complex Systems}}

\def\IJTP{{\sl Int.\ J. Theor.\ Phys.}}

\def\JPA{{\sl J. Phys.\ A:  Math.\ Gen.}}

\def\JSP{{\sl J. Stat.\ Phys.}}

\def\PD{{\sl Physica D}}

\def\PLA{{\sl Phys.\ Lett.\ A}}

\def\PRA{{\sl Phys.\ Rev.\ A}}

\def\PRD{{\sl Phys.\ Rev.\ D}}
\def\PRE{{\sl Phys.\ Rev.\ E}}

\def\SPJETP{{\sl Sov.\ Phys.\ JETP}}

\def\dajm{\hbox{D. A. Meyer}}

\def\rds{\hbox{R. D. Sorkin}}

\def\feynman{\hbox{R. P. Feynman}}

\def\gz{\hbox{G. Gr\"ossing and A. Zeilinger}}

\catcode`@=11
\newskip\ttglue

   \font\ninerm=cmr9    \font\eightrm=cmr8   \font\sixrm=cmr6
  \font\ninebf=cmbx9   \font\eightbf=cmbx8  \font\sixbf=cmbx6
  \font\nineit=cmti9   \font\eightit=cmti8  
  \font\ninesl=cmsl9   \font\eightsl=cmsl8  
  \font\ninemi=cmmi9   \font\eightmi=cmmi8  \font\sixmi=cmmi6

\font\bigtenbf=cmr10 scaled\magstep2 

\def\ninepoint{\def\rm{\fam0\ninerm}%
  \textfont0=\ninerm \scriptfont0=\sixrm
  \textfont1=\ninemi \scriptfont1=\sixmi
  \textfont\itfam=\nineit  \def\it{\fam\itfam\nineit}%
  \textfont\slfam=\ninesl  \def\sl{\fam\slfam\ninesl}%
  \textfont\bffam=\ninebf  \scriptfont\bffam=\sixbf
    \def\bf{\fam\bffam\ninebf}%
  \tt \ttglue=.5em plus.25em minus.15em
  \normalbaselineskip=11pt
  \setbox\strutbox=\hbox{\vrule height8pt depth3pt width0pt}%
  \normalbaselines\rm}

\def\eightpoint{\def\rm{\fam0\eightrm}%
  \textfont0=\eightrm \scriptfont0=\sixrm
  \textfont1=\eightmi \scriptfont1=\sixmi
  \textfont\itfam=\eightit  \def\it{\fam\itfam\eightit}%
  \textfont\slfam=\eightsl  \def\sl{\fam\slfam\eightsl}%
  \textfont\bffam=\eightbf  \scriptfont\bffam=\sixbf
    \def\bf{\fam\bffam\eightbf}%
  \tt \ttglue=.5em plus.25em minus.15em
  \normalbaselineskip=9pt
  \setbox\strutbox=\hbox{\vrule height7pt depth2pt width0pt}%
  \normalbaselines\rm}

\def\sfootnote#1{\edef\@sf{\spacefactor\the\spacefactor}#1\@sf
      \insert\footins\bgroup\eightpoint
      \interlinepenalty100 \let\par=\endgraf
        \leftskip=0pt \rightskip=0pt
        \splittopskip=10pt plus 1pt minus 1pt \floatingpenalty=20000
        \parskip=0pt\smallskip\item{#1}\bgroup\strut\aftergroup\@foot\let\next}
\skip\footins=12pt plus 2pt minus 2pt
\dimen\footins=30pc

\def\ie{{\it i.e.}}

\def\etc{{\it etc.}}

\def\Lemma{L{\eightpoint EMMA}}
\def\Theorem{T{\eightpoint HEOREM}}

\def\hfb{\hfil\break}

\magnification=1200
\input epsf.tex

\dimen0=\hsize \divide\dimen0 by 13 \dimendef\chasm=0
\dimen1=\hsize \advance\dimen1 by -\chasm \dimendef\usewidth=1
\dimen2=\usewidth \divide\dimen2 by 2 \dimendef\halfwidth=2
\dimen3=\usewidth \divide\dimen3 by 3 \dimendef\thirdwidth=3
\dimen4=\hsize \advance\dimen4 by -\halfwidth \dimendef\secondstart=4
\dimen5=\halfwidth \advance\dimen5 by -10pt \dimendef\indenthalfwidth=5
\dimen6=\thirdwidth \multiply\dimen6 by 2 \dimendef\twothirdswidth=6
\dimen7=\twothirdswidth \divide\dimen7 by 4 \dimendef\qttw=7
\dimen8=\qttw \divide\dimen8 by 4 \dimendef\qqttw=8
\dimen9=\qqttw \divide\dimen9 by 4 \dimendef\qqqttw=9

\parskip=0pt
\line{\hfil December 1995}
\line{\hfil {\it revised\/} September 1996}
\line{\hfil quant-ph/9604011}
\bigskip\bigskip\bigskip
\centerline{\bf\bigtenbf ON THE ABSENCE OF HOMOGENEOUS}
\bigskip
\centerline{\bf\bigtenbf SCALAR UNITARY CELLULAR AUTOMATA}
\vfill
\centerline{\bf David A. Meyer}
\bigskip 
\centerline{\sl Project in Geometry and Physics}
\centerline{\sl Department of Mathematics}
\centerline{\sl University of California/San Diego}
\centerline{\sl La Jolla, CA 92093-0112}
\centerline{dmeyer@euclid.ucsd.edu}
\vfill
\centerline{ABSTRACT}
\bigskip
%--------|---------|---------|---------|---------|---------|---------|
\noindent Failure to find homogeneous scalar unitary cellular automata
(CA) in one dimension led to consideration of only ``approximately 
unitary'' CA---which motivated our recent proof of a No-go Lemma in 
one dimension.  In this note we extend the one dimensional result to 
prove the absence of nontrivial homogeneous scalar unitary CA on 
Euclidean lattices in any dimension.
\bigskip
%--------|---------|---------|---------|---------|---------|---------|
\global\setbox1=\hbox{PACS numbers:\enspace}
\global\setbox2=\hbox{PACS numbers:}
\parindent=\wd1
\item{PACS numbers:}  05.30.-d,  % Quantum statistical mechanics
                      89.80.+h,  % Computer science and technology
                      03.65.Fd,  % Quantum mechanics.Algebraic methods 
                      11.30.-j.  % Symmetry and conservation laws
\item{\hbox to \wd2{KEY\hfill WORDS:}}   
                      quantum cellular automaton; quantum lattice gas;
                      unitarity;
\item{}               No-go Theorem.

\vfill
\centerline{to appear in \PLA.}
\vfill
\eject

\headline{\ninepoint\it Absence of homogeneous scalar unitary CA
          \hfil David A. Meyer}
\parskip=10pt
\parindent=20pt

%--------|---------|---------|---------|---------|---------|---------|
A classical cellular automaton (CA) consists of a lattice $L$ of 
{\sl cells\/} together with a {\sl field\/} 
$\phi : \N \times L \to S$, where $\N$ denotes the non-negative 
integers labelling timesteps and $S$ is the set of possible 
{\sl states\/} in which the field is valued.  Time evolution is 
locally defined; in the special case of an {\sl additive\/} CA the 
field evolves according to a {\sl local rule\/} of the form:
$$
\phi_{t+1}(x) = \sum_{e\in E(t,x)} w(t,x+e) \phi_t(x+e),      \eqno(1)
$$
where $E(t,x)$ is a set of lattice vectors defining local 
{\sl neighborhoods\/} for the automaton [1].  For the purposes of this
note, the lattice $L$ is taken to be generated by a set of $d$ 
linearly independent vectors in $\R^d$, \ie, as a group under vector
addition $L$ is isomorphic to $\Z^d$ or some periodic quotient 
thereof.  If $E(t,x)$ is a constant neighborhood and 
$w(t,x+e) \equiv w(x+e)$, the CA is {\sl homogeneous}.  For example, 
in the $\Z^2$ lattice generated by $v_1$ and $v_2$, the neighborhood
$E = \{0,\pm v_1,\pm v_2, \pm(v_1 + v_2)\}$ defines the ``triangular
lattice''.

%--------|---------|---------|---------|---------|---------|---------|
The additive evolution rule (1) is more compactly expressed as (left)
multiplication of the vector $\phi_t$ by an {\sl evolution\/} matrix 
having non-zero entries (the {\sl weights\/} $w(t,x+e)$) in row $x$ 
only in columns $x+e$ for $e \in E(t,x)$.  For example, if 
$S$ consists of the real numbers in the unit interval $[0,1]$, the 
weights $w(t,x+e)$ are positive, and the sum of the entries in each 
column of the evolution matrix is 1, then (1) defines a specific
{\sl probabilistic\/} CA [2].  The evolution preserves the $L^1$ norm 
of $\phi$:  $\sum_x \phi(x)$; if the $L^1$ norm of $\phi_0$ is one, 
then $\phi_t(x)$ may be interpreted physically as the probability that 
the system is in state $x$ at time $t$.  If the lattice of cells is a 
discretization of space, as suggested by the locality of the evolution 
rule (1), $\phi_t(x)$ is naturally interpreted to be the probability 
that a stochastic particle is in cell $x$ at time $t$.

%--------|---------|---------|---------|---------|---------|---------|
If the field is complex valued, or more precisely, if 
$S = \{z \in \C \mid |z| \le 1\}$, and the evolution matrix is 
{\sl unitary\/} then (1) defines what we refer to here as a 
{\sl scalar unitary\/} CA; this is a special case of a {\sl quantum\/}
CA (QCA) [3,4,5,6].  Unitary evolution preserves the $L^2$ norm of 
$\phi$:  $(\sum_x |\phi(x)|^2)^{1/2}$; if the $L^2$ norm of $\phi_0$ 
is 1, then $\phi_t(x)$ is the {\sl amplitude\/} for the system to be 
in, and $|\phi_t(x)|^2$ is the probability of observing, the state $x$ 
at time $t$.  Scalar QCA were first considered by Gr\"ossing and 
Zeilinger [4], although they found nontrivial homogeneous scalar CA in 
one dimension with neighborhoods of radius one (\ie, with the 
evolution matrix tridiagonal) only by relaxing their definition to 
allow ``approximately unitary'' evolution.  In [3] we showed that only 
trivial homogeneous scalar unitary CA exist in one dimension with 
neighborhoods of {\sl any\/} size:

%--------|---------|---------|---------|---------|---------|---------|
\noindent N{\eightpoint O-GO} \Lemma.  {\sl In one dimension there 
exists no nontrivial, homogeneous, scalar unitary CA.  More 
explicitly, every band $r$-diagonal unitary matrix which commutes with 
the 1-step translation matrix is also a translation matrix, times a 
phase.}

%--------|---------|---------|---------|---------|---------|---------|
The purpose of this note is to show that the analogous result also
holds in higher dimensions.  This will be important when we extend the 
one dimensional models of [3] to more realistic simulations of two or
three dimensional systems [7].  We shall give two different proofs of 
this No-go Theorem and then conclude by explaining how it may be 
evaded in order to find nontrivial QCA in any dimension.

%--------|---------|---------|---------|---------|---------|---------|
Consider first a lattice $L = \Z_{n_1} \oplus \cdots \oplus \Z_{n_d}$,
\ie, a finite lattice which is locally isomorphic to $\Z^d$ but is 
periodic in each coordinate.  The cells of this lattice may be ordered
lexicographically by their coordinates:  for a cell 
$(x_1,\ldots,x_d)$,
$$
x := x_d + n_d \, x_{d-1} + n_d n_{d-1} \, x_{d-2} + \cdots 
         + n_d \ldots n_2 \, x_1                              \eqno(2)
$$
defines the position of the cell in a one dimensional array.  In a CA 
with a neighborhood of radius $r$ the value of the field at this cell
depends on the values of the field at the cells 
$\{(y_1,\ldots,y_d) \mid |x_i - y_i| \le r\}$ at the previous 
timestep.  In the representation defined by (2), the evolution matrix
$U$ is what we may describe as ``depth $d$ band $r$-diagonal'' (the 
more familiar ``tridiagonal with fringes'' matrix arising in the 
finite difference method solution to a second order elliptic equation 
in two variables [8] is depth 2 band 1-diagonal in this terminology).  
More importantly for our purposes, $U$ is (sparsely) band 
$Kr$-diagonal, where
$$
K := 1 + n_d + n_d n_{d-1} + \cdots + n_d \ldots n_2,
$$
as shown in Figure 1.

%--------|---------|---------|---------|---------|---------|---------|
\midinsert
$$
\epsfxsize=\twothirdswidth\epsfbox{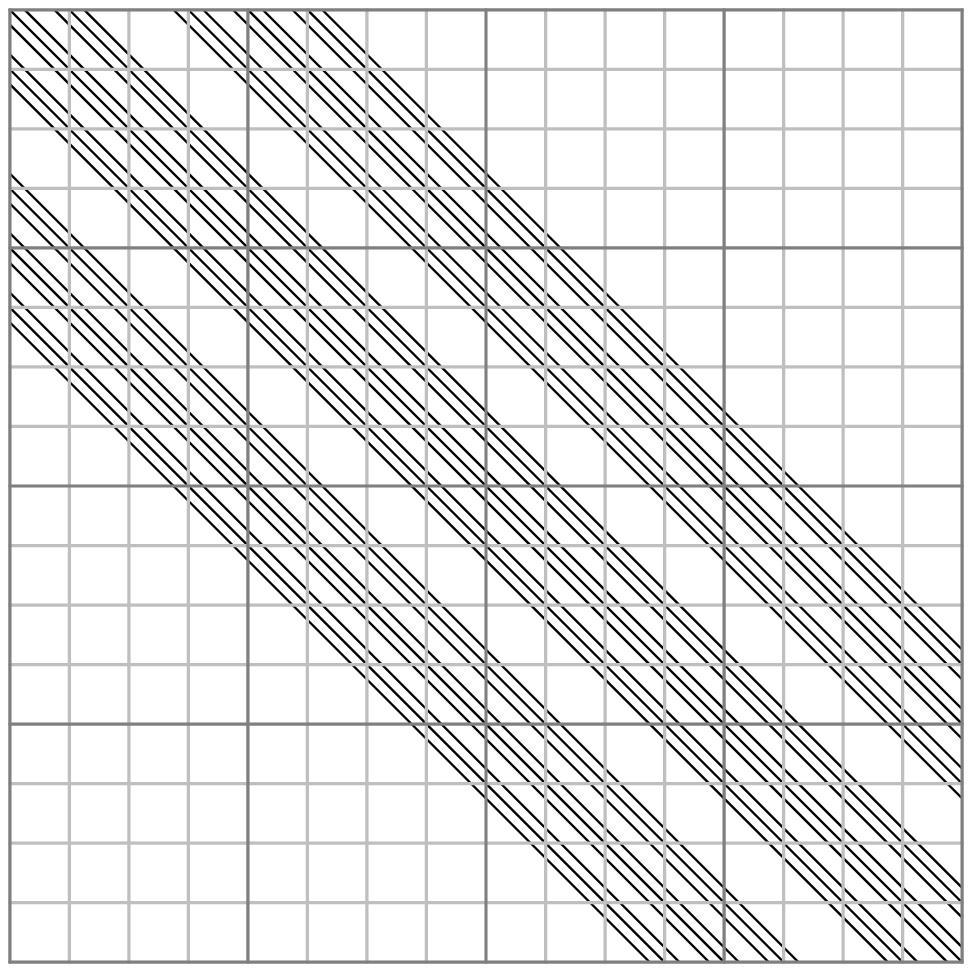}
$$
\eightpoint{%
{\narrower\noindent{\bf Figure 1}.  A portion of the depth 3 band 
$1$-diagonal evolution matrix $U$ for the lattice with dimensions 
$(n_1,n_2,n_3)$.  The small grey squares have size $n_3 \times n_3$; 
there are $n_2 \times n_2$ grey squares in each medium black square; 
and there are $n_1 \times n_1$ black squares in the whole array.  $U$ 
is band $1(1 + n_3 + n_3 n_2)$-diagonal.\par}
}
\endinsert

%--------|---------|---------|---------|---------|---------|---------|
The product of two band $Kr$-diagonal matrices is necessarily band
$2Kr$-diagonal.  The proof of the one dimensional No-go Lemma given in
[3] depends only on the size of $U$ being\break

%--------|---------|---------|---------|---------|---------|---------|
\moveright\secondstart\vtop to 0pt{\hsize=\halfwidth
\null\vbox to\vsize{\vfill
$$
\epsfxsize=\halfwidth\epsfbox{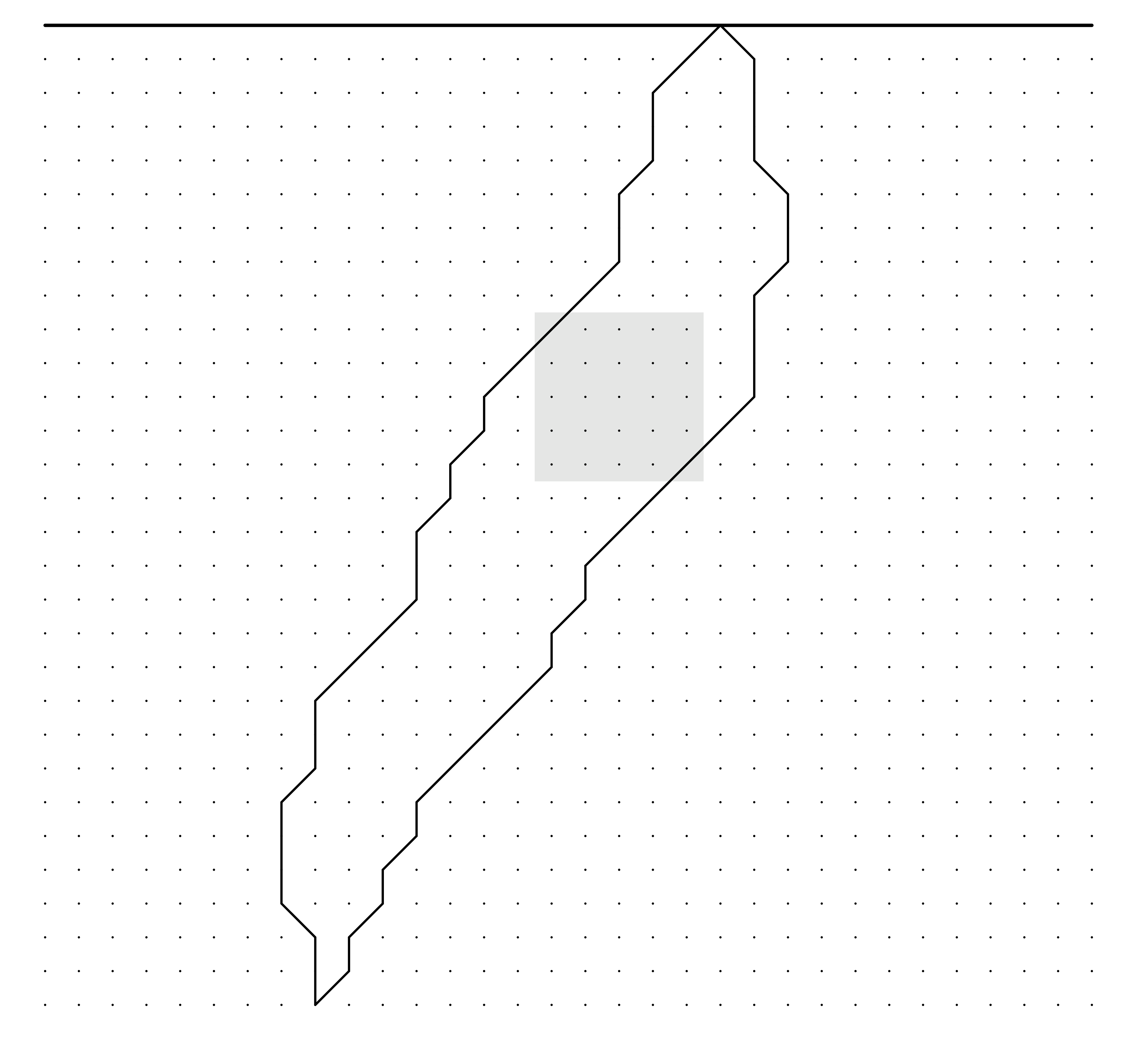}
$$
\eightpoint{%
\noindent{\bf Figure 2}.  A pair of spacetime histories of the quantum
particle in a one dimensional automaton with local neighborhood of 
radius 1 in the set $S$ defined by intersection with the shaded region 
$R$ of spacetime.  Since the histories coincide at the truncation time 
(which lies to the future of $R$), they contribute to the probability 
$|S|$.
\vskip 8.5pt
}}}
\vskip -2\baselineskip
\noindent large enough that the band $2Kr$-diagonal product 
$UU^{\dagger}$ is still band diagonal, namely that
$$
1 + 4Kr \le n_d \ldots n_1.                                   \eqno(3)
$$
Given any $r$, for a sufficiently large lattice $L$ 
(specifically, for sufficiently large $n_1$), inequality (3) is 
satisfied.

%--------|---------|---------|---------|---------|---------|---------|
The conclusion of the argument in [3] is that the only band diagonal
solution to $UU^{\dagger} = I$ is a phase times a matrix with only 
non-zero entries being ones along a single diagonal within the band.
This is a translation matrix even in the present higher dimensional
context.  Thus, as a scholium to the No-go Lemma for homogeneous 
scalar unitary CA in one dimension, we have proved:

%--------|---------|---------|---------|---------|---------|---------|
\noindent N{\eightpoint O-GO} \Theorem.  {\sl In any dimension the
only homogeneous, scalar unitary CA evolve by a constant translation 
with overall phase multiplication.}

\parshape=9
0pt \hsize
0pt \hsize
0pt \hsize
0pt \hsize
0pt \hsize
0pt \hsize
0pt \hsize
0pt \hsize
0pt \halfwidth
%--------|---------|---------|---------|---------|---------|---------|
Although the proof just given is straightforward, the physical and 
geometrical content of the result is perhaps obscured by the 
unraveling of the higher dimensional lattice $L$ into the one 
dimensional representation (2).  In fact, the theorem does not depend 
on the finiteness of the lattice which was necessary for the band 
diagonality of the $U$ shown in Figure 1.  To rectify this problem let 
us consider a second argument using a sum-over-histories approach.  In 
[3] we saw that this is particularly natural since a scalar QCA may be 
interpreted to be a quantum particle automaton:  the system consists 
of a single particle moving on the lattice, $\phi_t(x)$ is the 
amplitude for the particle to be in state $x$ at time $t$, and the 
weight $w(t,x+e)$ is the amplitude for the particle to move from $x+e$ 
to $x$.

\parshape=1
0pt \halfwidth
%--------|---------|---------|---------|---------|---------|---------|
In the sum-over-histories framework for quantum mechanics a 
probability is associated to a set $S$ of particle histories (defined 
by boolean expressions in projectors onto states $x_i$ at times $t_i$) 
by the rule:
$$
|S| = \sum_{\gamma_1,\gamma_2 \in S} 
       w(\gamma_1) \overline w(\gamma_2) 
       \delta\bigl(\gamma_1(T),\gamma_2(T)\bigr),
$$
where the delta function ensures that the only non-zero contributions
to the probability come from pairs of paths in $S$ which coincide at
the {\sl truncation time\/} $T$ [9].  Of course, as shown in Figure 2, 
only truncation times defining spacelike hypersurfaces entirely to the 
future of the conditions defining $S$ are permitted.  {\sl Unitarity 
is the invariance of probability under a change in truncation 
time.}  ~~That is, for 
\eject
\noindent any two states $x_1$ and $x_2$, the sum of the 
contributions of all pairs of paths, one from each of these states at
time $T_1$ to any common state $x$ at time $T_2 > T_1$, must vanish 
unless $x_1 \equiv x_2$, in which case it must be one.

%--------|---------|---------|---------|---------|---------|---------|
In particular, this condition applies to the paths corresponding to
advancing the truncation time by one timestep.  In a homogeneous CA,
a cell $x_1$ may influence cells in the constant neighborhood $E$ 
around it at the next timestep; hence any pair of paths, one from each 
of $x_1$ and $x_2$, which coincide at the next timestep, do so at a 
cell in the intersection of the neighborhood around $x_1$ and the 
neighborhood around $x_2$.  The unitarity conditions on the weights in
(1) thus arise from each pair of cells with intersecting 
neighborhoods:  the corresponding sum vanishes except when the two
cells coincide, in which case it is one.

%--------|---------|---------|---------|---------|---------|---------|
With this description of the unitarity conditions it is easy to prove
the No-go Theorem.  Order the cells in the neighborhood of $x_1$ as in
(2).  Let $k$ be the position of the first non-zero weight $w_k$ in 
this ordering (there must be one since the zero matrix is not 
unitary) and let $e_k$ denote the corresponding lattice vector.  
Consider $x_2 := x_1 + e_{|E|} - e_k$.  The set of cells with possibly 
non-zero weights in the neighborhood of $x_2$ intersects the 
neighborhood of $x_1$ only at the last cell in that ordering, so 
$w_k \overline w_{|E|} = 0$ and hence $w_{|E|} = 0$.  Now slide the
second neighborhood down one as in Figure 3, \ie, let 
$x_2 := x_1 + e_{|E|-1} - e_k$.  The set of cells with possibly 
non-zero weights in the intersection of $x_1$ and $x_2$ is again a 
singleton, still labelled $k$ in the neighborhood of $x_2$, but now 
$|E|-1$ in the neighborhood of $x_1$.  So 
$w_k \overline w_{|E|-1} = 0$ and hence $w_{|E|-1} = 0$.  Continue
this process until $x_2 = x_1$, whence unitarity requires
$w_k \overline w_k = 1$.  The conclusion of the No-go Theorem follows:
the one step evolution of this homogeneous scalar unitary CA is 
translation by $e_k$ and multiplication by the phase $w_k$.

%--------|---------|---------|---------|---------|---------|---------|
\midinsert
$$
\epsfxsize=\usewidth\epsfbox{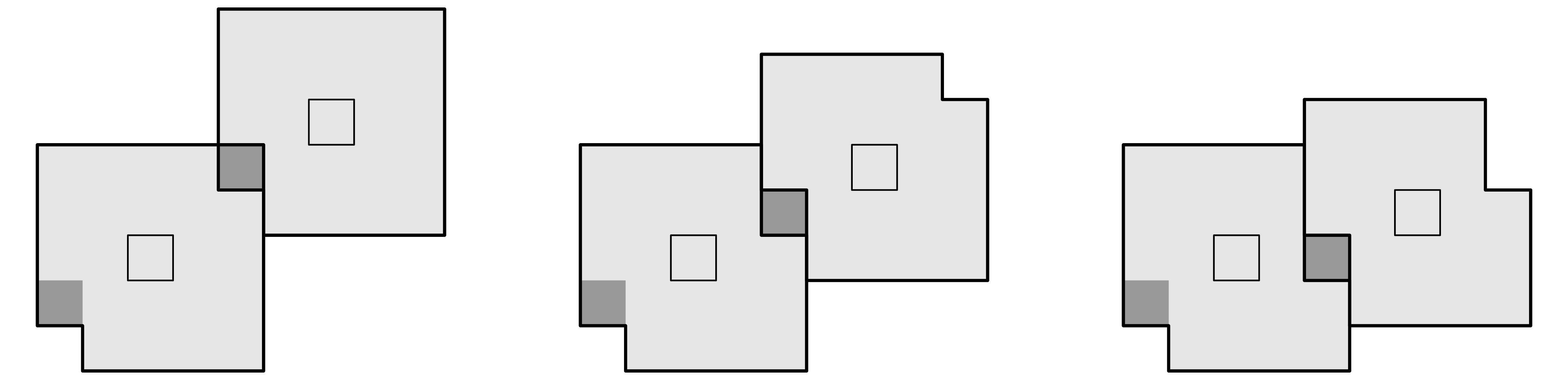}
$$
\eightpoint{%
{\narrower\noindent{\bf Figure 3}.  Intersecting neighborhoods of 
radius 2 in a two dimensional CA; in each pair $x_1$ is the lower left 
cell outlined and $x_2$ is the upper right one.  The first non-zero 
weight is at position 2 (dark grey) in each neighborhood.  In the 
three steps of the argument shown, $x_2$ is shifted so that the cell 
at position 2 in its neighborhood coincides successively with the cell 
at position $|E| = 25$, then 24, then 23 in the neighborhood of $x_1$.  
When the last column has been depleted the process is repeated on the 
next to last, \etc, until the neighborhoods coincide.\par} 
}
\endinsert

%--------|---------|---------|---------|---------|---------|---------|
The homogeneity hypothesis in the No-go Theorem is the requirement 
that the evolution matrix be invariant under the action of the 
translation group of the lattice.  The conclusion is that this 
restriction on scalar unitary CA renders them too simple to be of much 
interest.  As we showed in [3], however, if the evolution matrix is 
required to be invariant only under the action of a subgroup of the 
translation group of the one dimensional lattice, the No-go Lemma is 
evaded and there are many interesting scalar QCA (the first of which 
seems to have been described by Feynman [10]; similar discrete models 
for a quantum particle have been studied by several authors more 
recently [5,11]).  This is equally true in higher dimensional 
lattices:  the one step evolution of a quantum {\sl partitioning\/} 
[2,12] CA is invariant under the action of a subgroup of the 
translations on the lattice and may be interpreted to be composed of 
particle scattering matrices.  Higher dimensional quantum particle 
automata [7,13] and their generalizations to quantum lattice gas 
automata [7,14] have been constructed in exactly this way.
\bigskip

\global\setbox1=\hbox{[00]\enspace}
\parindent=\wd1

\noindent{\bf References}
\medskip

\parskip=0pt
%--------|---------|---------|---------|---------|---------|---------|
\item{[1]}
S. Wolfram,
``Computation theory of cellular automata'',
\CMP\ {\bf 96} (1984) 15--57.

\item{[2]}
P. Ruj\'an,
``Cellular automata and statistical mechanical models'',
\JSP\ {\bf 49} (1987) 139--222;\hfb
A. Georges and P. Le Doussal,
``From equilibrium spin models to probabilistic cellular automata'',
\JSP\ {\bf 54} (1989) 1011--1064;\hfb
and references therein.

\item{[3]}
\dajm,
``From quantum cellular automata to quantum lattice gases'',
UCSD preprint (1995), quant-ph/9604003, to appear in \JSP

\item{[4]}
\gz,
``Quantum cellular automata'',
\CS\ {\bf 2}\break
(1988) 197--208;\hfb
\gz,
``A conservation law in quantum cellular automata'',
\PD\ {\bf 31} (1988) 70--77;\hfb
S. Fussy, G. Gr\"ossing, H. Schwabl and A. Scrinzi,
``Nonlocal computation in quantum cellular automata'',
\PRA\ {\bf 48} (1993) 3470--3477.

\item{[5]}
I. Bialynicki-Birula,
``Weyl, Dirac, and Maxwell equations on a lattice as unitary 
  cellular automata'',
\PRD\ {\bf 49} (1994) 6920--6927.

\item{[6]}
C. D\"urr, H. L\^e Thanh and M. Santha,
``A decision procedure for well-formed linear quantum cellular 
  automata'',
in C. Puecha and R. Reischuk, eds.,
{\sl STACS 96:  Proceedings of the 13th Annual Symposium on 
     Theoretical Aspects of Computer Science},
Grenoble, France, 22--24 February 1996,
{\sl Lecture notes in computer science} {\bf 1046}
(New York:  Springer-Verlag 1996) 281--292;\hfb
C. D\"urr and M. Santha,
``A decision procedure for unitary linear quantum cellular
automata'',
preprint (1996) quant-ph/9604007;\hfb
\dajm,
``Unitarity in one dimensional nonlinear quantum cellular automata'',
UCSD preprint (1996), quant-ph/9605023;\hfb
W. van Dam,
``A universal quantum cellular automaton'',
preprint (1996).

\item{[7]}
\dajm,
in preparation.

\item{[8]}
W. H. Press, B. P. Flannery, S. A. Teukolsky and W. T. Vetterling,
{\sl Numerical Recipes:  The art of scientific computing\/}
(Cambridge: Cambridge University Press 1986) 619--620.

\item{[9]}
\rds,
``On the role of time in the sum-over-histories framework for 
  gravity'', 
presented at the conference on The History of Modern Gauge 
  Theories, Logan, Utah, July 1987,
published in \IJTP\ {\bf 33} (1994) 523--534;\hfb
\rds,
``Problems with causality in the sum-over-histories framework for
  quantum mechanics'',
in A. Ashtekar and J. Stachel, eds.,
{\sl Conceptual Problems of Quantum Gravity}, proceedings of the
  Osgood Hill Conference, North Andover, MA, 15--19 May 1988
(Boston:  Birkh\"auser 1991) 217--227;\hfb
J. B. Hartle,
``The quantum mechanics of closed systems'',
in B.-L. Hu, M. P. Ryan and C. V. Vishveshwara, eds.,
{\sl Directions in General Relativity:  Proceedings of the 1993
  international symposium, Maryland.  Volume 1:  papers in honor
  of Charles Misner\/}
(Cambridge: Cambridge University Press 1993) 104--124;\hfb
and references therein.

\item{[10]}
\feynman\ and A. R. Hibbs,
{\sl Quantum Mechanics and Path Integrals}
(New York:  McGraw-Hill 1965).

\item{[11]}
S. Succi and R. Benzi,
``Lattice Boltzmann equation for quantum mechanics'',
\PD\ {\bf 69} (1993) 327--332;\hfb
S. Succi,
``Numerical solution of the the Schr\"odinger equation using 
  discrete kinetic theory'',
\PRE\ {\bf 53} (1996) 1969--1975;\hfb
M. D. Kostin,
``Cellular automata for quantum systems'',
\JPA\ {\bf 26} (1993) L209--L215.

\item{[12]}
T. Toffoli and N. H. Margolus,
``Invertible cellular automata:  a review'',
\PD\ {\bf 45} (1990) 229--253.

\item{[13]}
G. V. Riazanov,
``The Feynman path integral for the Dirac equation'',
\SPJETP\ {\bf 6} (1958) 1107--1113.

\item{[14]}
B. M. Boghosian and W. Taylor, IV,
``A quantum lattice-gas model for the many-particle Schr\"odinger
  equation in $d$ dimensions'',
preprint (1996) BU-CCS-960401, PUPT-1615, quant-ph/9604035.

\bye